\documentstyle[psfig]{elsart}

\let\text\mbox

\begin{document}

\begin{frontmatter}

\title{Search for CP Violation in Charged D Meson Decays}

\collab{Fermilab E791 Collaboration}

\author[inst_9]{E.~M.~Aitala,}
\author[inst_1]{S.~Amato,}
\author[inst_1]{J.~C.~Anjos,}
\author[inst_5]{J.~A.~Appel,}
\author[inst_15]{D.~Ashery,}
\author[inst_5]{S.~Banerjee,}
\author[inst_1]{I.~Bediaga,}
\author[inst_8]{G.~Blaylock,}
\author[inst_16]{S.~B.~Bracker,}
\author[inst_14]{P.~R.~Burchat,}
\author[inst_6]{R.~A.~Burnstein,}
\author[inst_5]{T.~Carter,}
\author[inst_1]{H.~S.~Carvalho,}
\author[inst_13]{N.~K.~Copty,}
\author[inst_9]{L.~M.~Cremaldi,}
\author[inst_19]{C.~Darling,}
\author[inst_5]{K.~Denisenko,}
\author[inst_12]{A.~Fernandez,}
\author[inst_2]{P.~Gagnon,}
\author[inst_1]{C.~Gobel,}
\author[inst_9]{K.~Gounder,}
\author[inst_5]{A.~M.~Halling,}
\author[inst_4]{G.~Herrera,}
\author[inst_15]{G.~Hurvits,}
\author[inst_5]{C.~James,}
\author[inst_6]{P.~A.~Kasper,}
\author[inst_5]{S.~Kwan,}
\author[inst_13]{D.~C.~Langs,}
\author[inst_2]{J.~Leslie,}
\author[inst_5]{B.~Lundberg,}
\author[inst_15]{S.~MayTal-Beck,}
\author[inst_3]{B.~Meadows,}
\author[inst_1]{J.~R.~T.~de~Mello~Neto,}
\author[inst_17]{R.~H.~Milburn,}
\author[inst_1]{J.~M.~de~Miranda,}
\author[inst_17]{A.~Napier,}
\author[inst_7]{A.~Nguyen,}
\author[inst_3,inst_12]{A.~B.~d'Oliveira,}
\author[inst_2]{K.~O'Shaughnessy,}
\author[inst_6]{K.~C.~Peng,}
\author[inst_3]{L.~P.~Perera,}
\author[inst_13]{M.~V.~Purohit,}
\author[inst_9]{B.~Quinn,}
\author[inst_18]{S.~Radeztsky,}
\author[inst_9]{A.~Rafatian,}
\author[inst_7]{N.~W.~Reay,}
\author[inst_9]{J.~J.~Reidy,}
\author[inst_1]{A.~C.~dos Reis,}
\author[inst_6]{H.~A.~Rubin,}
\author[inst_3]{A.~K.~S.~Santha,}
\author[inst_1]{A.~F.~S.~Santoro,}
\author[inst_11]{A.~J.~Schwartz,}
\author[inst_18]{M.~Sheaff,}
\author[inst_7]{R.~A.~Sidwell,}
\author[inst_19]{A.~J.~Slaughter,}
\author[inst_3]{M.~D.~Sokoloff,}
\author[inst_7]{N.~R.~Stanton,}
\author[inst_18]{K.~Stenson,}
\author[inst_9]{D.~J.~Summers,}
\author[inst_19]{S.~Takach,}
\author[inst_5]{K.~Thorne,}
\author[inst_10]{A.~K.~Tripathi,}
\author[inst_18]{S.~Watanabe,}
\author[inst_15]{R.~Weiss-Babai,}
\author[inst_11]{J.~Wiener,}
\author[inst_7]{N.~Witchey,}
\author[inst_19]{E.~Wolin,}
\author[inst_9]{D.~Yi,}
\author[inst_7]{S. Yoshida,}                         
\author[inst_14]{R.~Zaliznyak,}
\author[inst_7]{C.~Zhang}

\address[inst_1]{Centro Brasileiro de Pesquisas F{\'i}sicas, Rio de Janeiro, 
                 Brazil}
\address[inst_2]{University of California, Santa Cruz, California 95064}
\address[inst_3]{University of Cincinnati, Cincinnati, Ohio 45221}
\address[inst_4]{CINVESTAV, Mexico}
\address[inst_5]{Fermilab, Batavia, Illinois 60510}
\address[inst_6]{Illinois Institute of Technology, Chicago, Illinois 60616}
\address[inst_7]{Kansas State University, Manhattan, Kansas 66506}
\address[inst_8]{University of Massachusetts, Amherst, Massachusetts 01003}
\address[inst_9]{University of Mississippi, University, Mississippi 38677}
\address[inst_10]{The Ohio State University, Columbus, Ohio 43210}
\address[inst_11]{Princeton University, Princeton, New Jersey 08544}
\address[inst_12]{Universidad Autonoma de Puebla, Mexico}
\address[inst_13]{University of South Carolina, Columbia, South Carolina 29208}
\address[inst_14]{Stanford University, Stanford, California 94305}
\address[inst_15]{Tel Aviv University, Tel Aviv, Israel}
\address[inst_16]{317 Belsize Drive, Toronto, Canada}
\address[inst_17]{Tufts University, Medford, Massachusetts 02155}
\address[inst_18]{University of Wisconsin, Madison, Wisconsin 53706}
\address[inst_19]{Yale University, New Haven, Connecticut 06511}

%%%%%%

\begin{abstract} 
We report results of a search for CP violation in the singly 
Cabibbo-suppressed decays $D^+ \rightarrow K^- K^+ \pi^+$, 
$\phi \pi^+$, $\overline{K}^*(892)^0 K^+$, and $\pi^- \pi^+ \pi^+$
based on data from the charm hadroproduction experiment E791 at Fermilab. 
We search for a difference in the $D^+$ and $D^-$ decay rates for each 
of the final states.  No evidence for a difference is seen.  
The decay rate asymmetry parameters ($A_{CP}$), defined as the difference 
in the $D^+$ and $D^-$ decay rates divided by the sum of the decay rates, 
are measured to be:
$A_{CP}(KK\pi)     = -0.014 \pm 0.029$, 
$A_{CP}(\phi\pi)   = -0.028 \pm 0.036$, 
$A_{CP}(K^*(892)K) = -0.010 \pm 0.050$, and
$A_{CP}(\pi\pi\pi) = -0.017 \pm 0.042$.


\end{abstract}

\end{frontmatter}

%%%%%%

CP violation can be accommodated in the Standard Model (SM)
by a complex phase in the Cabibbo-Kobayashi-Maskawa matrix describing the
transitions between quarks induced by the charged weak interaction.
To date, CP violation has been observed only in the neutral kaon
system.  It is being actively searched for in $B$ decays, where the
effects of CP violation in the SM are expected to be
large~\cite{ref_bot}, and in hyperon decays~\cite{ref_hyp}. 
In contrast to the strange and bottom sectors, the SM predictions
of CP violation for charm 
decays are much smaller~\cite{ref_sm}, making the charm sector a good
place to test the SM and to look for evidence of physics 
beyond the SM~\cite{ref_bsm}.

CP violation occurs if the decay rate for a particle differs from its CP
conjugate decay rate:
$\Gamma(D\rightarrow f) \neq \Gamma(\overline{D}\rightarrow \overline{f})$. 
Such an asymmetry requires the interference of at least two independent
amplitudes with non-zero relative phase. In $D$
decays, as in $K$ and $B$ decays, there are two classes of CP
violation: indirect and direct. In the case of indirect CP violation,
the asymmetry is associated with mixing, which can occur only in
neutral meson decays.
In the case of direct CP violation, final state interactions must 
induce a strong phase shift and both the strong and weak phases of 
the two amplitudes must differ. This can occur in both
charged and neutral meson decays.
In the SM, direct CP-violation asymmetries in $D$ decays 
are predicted to be largest in singly Cabibbo-suppressed (SCS) decays 
(at most of the order of $10^{-3}$)
and non-existent in Cabibbo-favored (CF) and doubly 
Cabibbo-suppressed (DCS) decays~\cite{ref_sm}.

Current experimental sensitivity to decay rate asymmetries due to direct CP
violation, defined below, is of the order of $10^{-1}$.
Experimental results on searches for CP violation in $D^0$ decays come from
E691~\cite{ref_e691}, E687~\cite{ref_e687}, and
CLEO~\cite{ref_cleo}. However, limits on CP violation in $D^+$ decays
come only from E687~\cite{ref_e687}.
In this letter, we report on a higher-statistics search for evidence 
of direct CP violation in the SCS decays $D^+ \rightarrow K^- K^+ \pi^+$ 
(inclusive), 
$\phi \pi^+$, $\overline{K}^*(892)^0 K^+$, and, for the first time,
$\pi^- \pi^+ \pi^+$. 

The signature for CP violation in charged D decays 
is an asymmetry in the decay rates:
\begin{equation}
A_{CP} = \frac{\Gamma(D^+ \rightarrow f^+) - \Gamma(D^- \rightarrow f^-)}
              {\Gamma(D^+ \rightarrow f^+) + \Gamma(D^- \rightarrow f^-)}.
\label{eq_1}
\end{equation}
To the extent that
\begin{equation}
\frac{\epsilon(D^+ \rightarrow f^+_{SCS})}
     {\epsilon(D^+ \rightarrow K^- \pi^+ \pi^+)} = 
\frac{\epsilon(D^- \rightarrow f^-_{SCS})}
     {\epsilon(D^- \rightarrow K^+ \pi^- \pi^-)},
\label{eq_eff} 
\end{equation}
where $\epsilon$ is our detector efficiency and 
$f^\pm_{SCS}$ is $K^\mp K^\pm \pi^\pm$, $\phi \pi^\pm$,  
$K^*(892) K^\pm$, or $\pi^\mp \pi^\pm \pi^\pm$,
we can normalize the production rates relative to the
CF mode $D^\pm \rightarrow K^\mp \pi^\pm \pi^\pm$.
Then, Eq.~(\ref{eq_1}) becomes
\begin{equation}
A_{CP} = \frac{\eta(D^+) - \eta(D^-)}{\eta(D^+) + \eta(D^-)},
\label{eq_2}
\end{equation}
where
\begin{equation}
\eta(D^\pm) = \frac{N(D^\pm \rightarrow f^\pm_{SCS})}
                 {N(D^\pm \rightarrow K^\mp \pi^\pm \pi^\pm)}
\label{eq_3}
\end{equation}
and $N$ is the number of observed $D$ decays.
Thus, differences in the $D^\pm$ production rates~\cite{ref_asym},  
relative efficiencies, 
and most other sources of systematic errors cancel.

E791 is a high statistics charm experiment
which ran at Fermilab during the 1991-1992 fixed-target run~\cite{ref_detect}.
The experiment combined an extremely
fast data acquisition system with an open trigger to record
the world's largest sample of open charm.
Over 20 billion events were collected using the Tagged Photon Spectrometer
with a 500~GeV $\pi^-$ beam. There were five target foils: one 0.5~mm thick 
platinum foil followed by four 1.6~mm thick diamond foils with 15~mm 
center-to-center separations. The spectrometer included
23 planes of silicon microstrip detectors (6 upstream and
17 downstream of the target), 2 dipole magnets, 10 planes of proportional wire 
chambers (8 upstream and 2 downstream of the target),
35 drift chamber planes, 2 multicell threshold \v{C}erenkov counters
that provided $\pi/K$ separation in the 6--60~GeV/$c$ momentum 
range~\cite{ref_ckv}, electromagnetic and hadronic calorimeters, and a
muon detector.

All selection criteria (cuts) used to choose candidate SCS 
decays $D^\pm \rightarrow K^\mp K^\pm \pi^\pm$, $\phi \pi^\pm$, 
$K^*(892) K^\pm$, and $\pi^\mp \pi^\pm \pi^\pm$, 
with the exception of \v{C}erenkov identification requirements,  
were optimized 
to maximize $S/\sqrt{S + B}$, where $S$ is the number of
CF $D^\pm \rightarrow K^\mp \pi^\pm \pi^\pm$  signal events, 
scaled to the level expected for each SCS decay mode,
and $B$ is the the number of background events for each of the signals studied.
To reconstruct the $D^\pm$ candidates, three-prong decay vertices with 
charge of $\pm 1$ were selected. 
All decay tracks were required to travel through at least one magnet
and be of good quality.
Decay vertices were required to be located outside the target foils,
and the significance of separation from the primary vertex in the 
beam direction ($\Delta z/\sigma_{\Delta z}$)
was required to be at least 11 for $KK\pi$,
8 for $\phi\pi$, 9 for $K^*K$, and 14 for $\pi\pi\pi$. 
The component of the $D$ momentum perpendicular to the line 
joining the primary and secondary vertices was required to be less 
than 0.35~GeV/$c$ for $KK\pi$, $\phi\pi$, and $K^*K$, and  
0.20~GeV/$c$ for $\pi\pi\pi$. The impact parameter of the $D$ 
momentum with respect to the primary vertex was required to be less than
55~$\mu$m for $KK\pi$ and $K^*K$, 80~$\mu$m for $\phi\pi$, and 40~$\mu$m
for $\pi\pi\pi$. Decay tracks were required to  pass closer 
to the secondary than to the primary vertex and the sum of the square of 
their momenta, perpendicular to the $D$ direction, 
was required to be larger than 0.30~(GeV/$c$)$^2$ for $KK\pi$, $K^*K$, 
and $\pi\pi\pi$, and 0.15~(GeV/$c$)$^2$ for $\phi\pi$. 

In the decay $D^\pm \rightarrow K^\mp \pi^\pm \pi^\pm$,
the kaon was identified on the basis of charge alone
and no \v{C}erenkov identification cuts were applied.
Also, in the decay $D^\pm \rightarrow \phi \pi^\pm$, 
no \v{C}erenkov identification cuts were 
applied due to the $\phi$ mass selection criteria.
In the decays $D^\pm \rightarrow K^\mp K^\pm \pi^\pm$ and $K^*(892) K^\pm$,
\v{C}erenkov identification cuts improved the signal significance. 
In these decay modes, tracks considered as kaons were required 
to have kaon signatures in the \v{C}erenkov counters.
No \v{C}erenkov identification cuts were applied for the decay 
$D^\pm \rightarrow \pi^\mp \pi^\pm \pi^\pm$.
To minimize the systematic uncertainties 
when extracting $A_{CP}$, 
we applied the same cuts, with the exception of \v{C}erenkov identification, 
to the SCS decays and to the normalizing CF decay
for each of the decay modes studied.

The mean momentum of surviving $D^\pm$ mesons was 70~GeV/$c$. Longitudinal 
and transverse position resolutions for the primary vertex were 350~$\mu$m
and 6~$\mu$m, respectively. For secondary vertices from $D^\pm$  decays,
the transverse resolution was about 9~$\mu$m, nearly independent of the 
$D^\pm$ momentum, and the longitudinal resolution was about 360~$\mu$m at 
70~GeV/$c$ and increased by 30~$\mu$m every 10~GeV/$c$.

\begin{figure}
\vspace*{2 cm}
\centerline{\psfig{figure=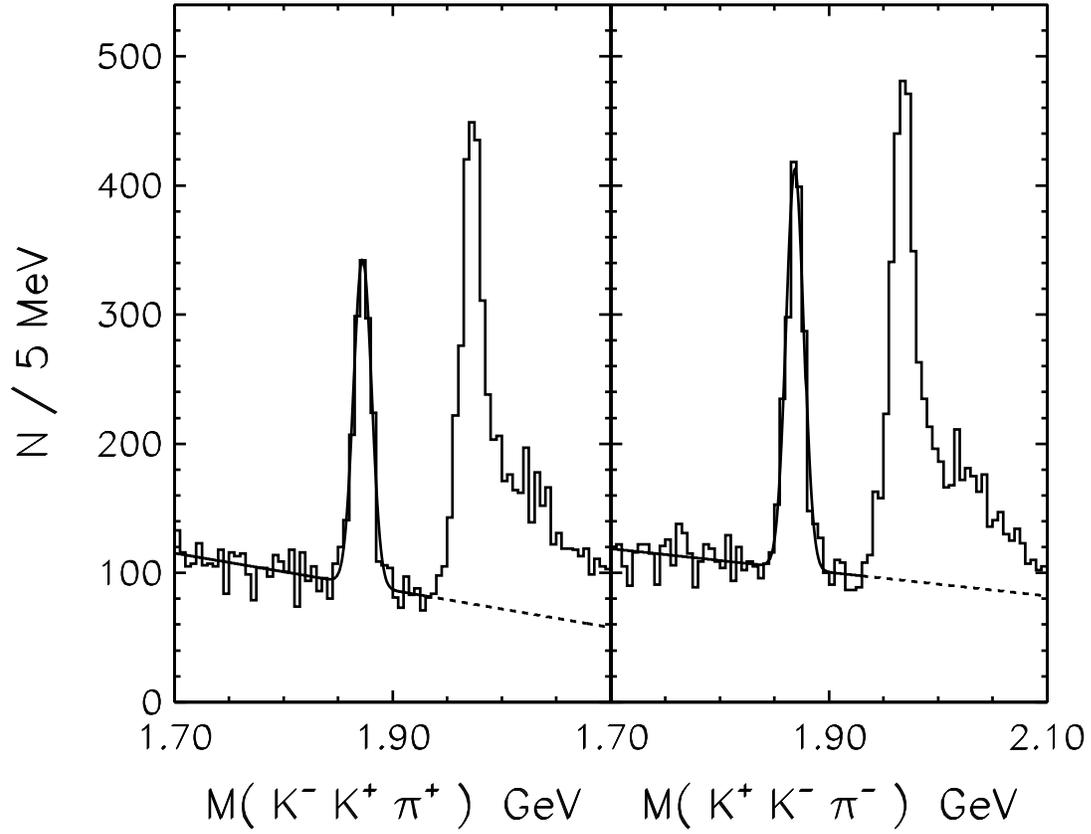,height=4.5in}}
\vspace*{3 cm}
\caption[]{The fits of $K^- K^+ \pi^+$ and $K^+ K^- \pi^-$ mass distributions,
as described in the text.
The $D_s^\pm$ and $K^\mp \pi^\pm \pi^\pm$ reflection region above 1.93~GeV 
was excluded from both fits.
The dashed lines represent the fits extended into the excluded regions.}
\vspace*{2 cm}
\label{fig_kkpi}
\end{figure}

\begin{figure}
\vspace*{2 cm}
\centerline{\psfig{figure=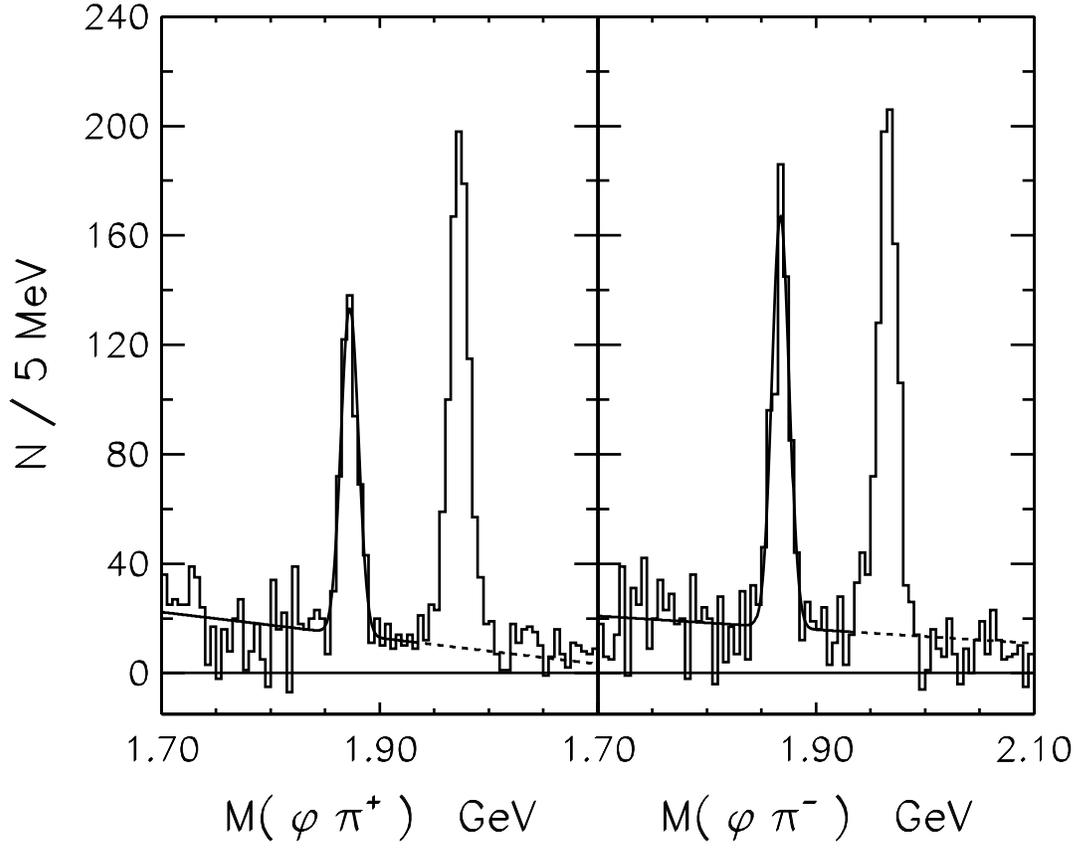,height=4.5in}}
\vspace*{3 cm}
\caption[]{The fits of $\phi \pi^+$ and $\phi \pi^-$ mass distributions,
as described in the text.
The $D_s^\pm$ 
region above 1.93~GeV was excluded from both fits.
The dashed  lines represent the fits extended into the excluded regions.}
\vspace*{2 cm}
\label{fig_phi}
\end{figure}

\begin{figure}
\vspace*{2 cm}
\centerline{\psfig{figure=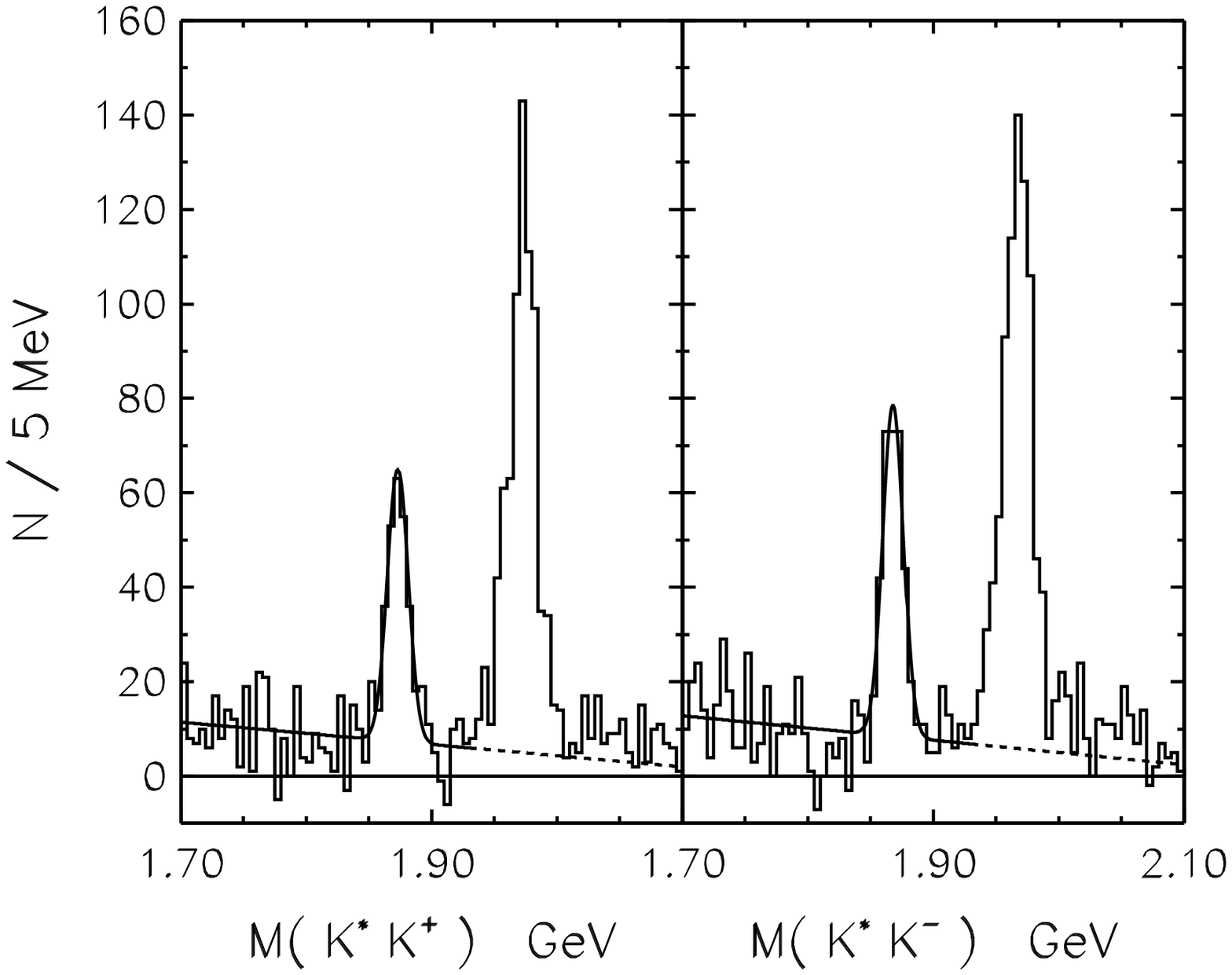,height=4.5in}}
\vspace*{3 cm}
\caption[]{The fits of $\overline{K}^*(892)^0 K^+$ and $K^*(892)^0 K^-$  
mass distributions,
as described in the text.
The $D_s^\pm$ and $K^\mp \pi^\pm \pi^\pm$ reflection region above 1.93~GeV 
was excluded from both fits.
The dashed lines represent the fits extended into the excluded regions.}
\vspace*{2 cm}
\label{fig_kst}
\end{figure}

\begin{figure}                                              
\vspace*{2 cm}
\centerline{\psfig{figure=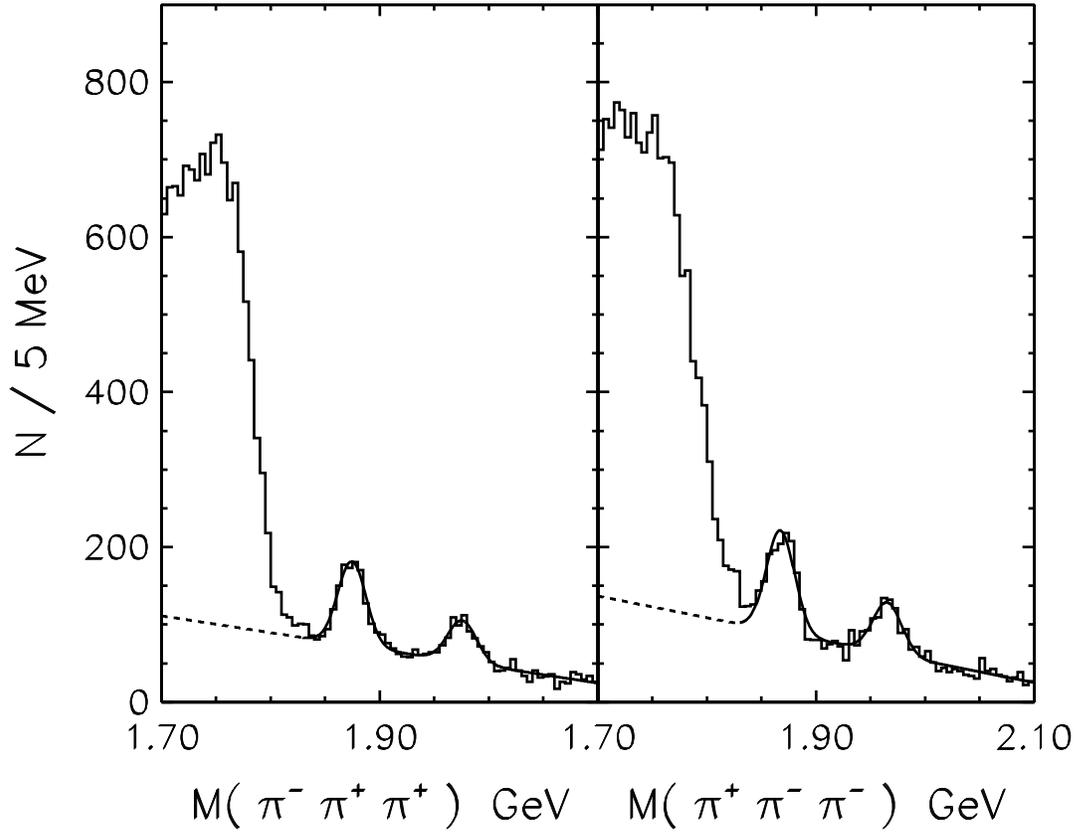,height=4.5in}}
\vspace*{3 cm}
\caption[]{The fits of $\pi^- \pi^+ \pi^+$ and $\pi^+ \pi^- \pi^-$ 
mass distributions, as described in the text. 
The $K^\mp \pi^\pm \pi^\pm$ reflection
region below 1.83~GeV was excluded from both fits.
The dashed lines represent the fits extended into the excluded regions.}
\vspace*{2 cm}
\label{fig_3pi}
\end{figure}

\begin{figure}
\vspace*{2 cm}
\centerline{\psfig{figure=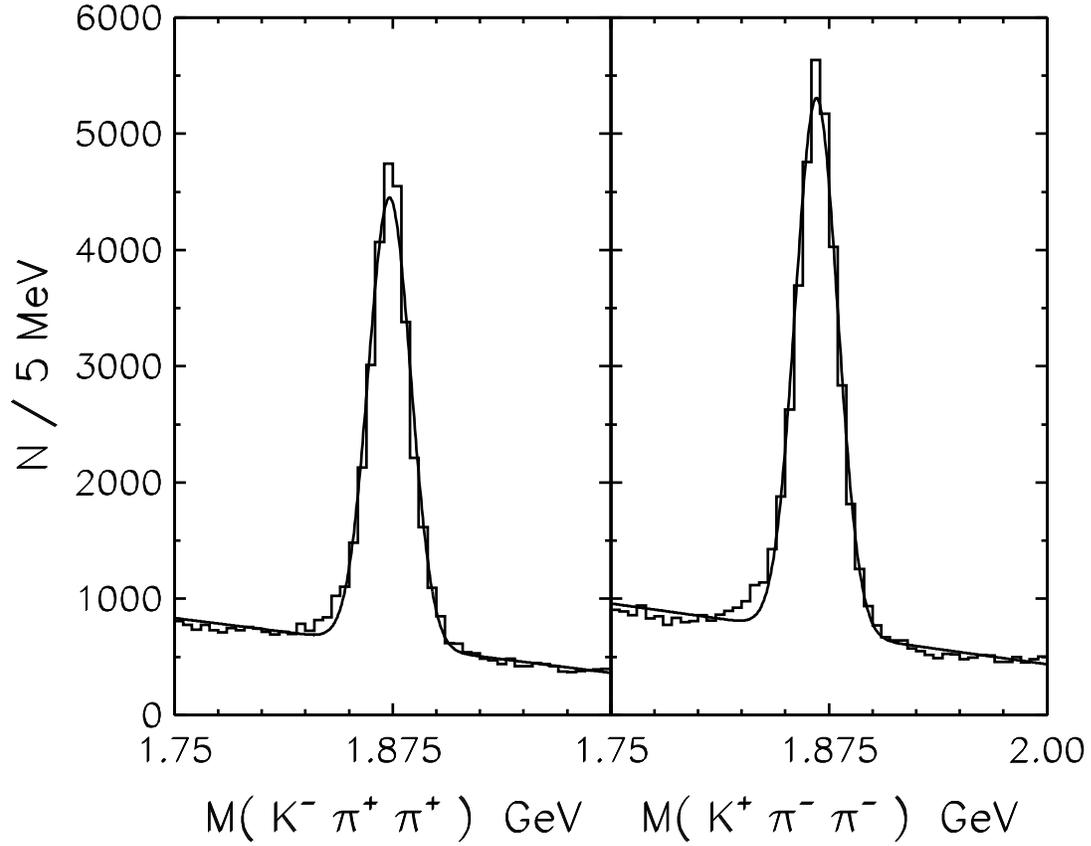,height=4.5in}}
\vspace*{3 cm}
\caption[]{The fits of the $K^- \pi^+ \pi^+$ and $K^+ \pi^- \pi^-$ 
mass distributions for $K^\mp \pi^\pm \pi^\mp$ candidates that survive 
the $D^\pm \rightarrow K^\mp K^\pm \pi^\pm$ 
selection criteria, as described in the text.}
\vspace*{2 cm}
\label{fig_kpipi}
\end{figure}

To determine the yields for each decay mode, we
performed simultaneous binned maximum likelihood fits  
of the mass distributions (see  Figs.~\ref{fig_kkpi}--\ref{fig_kpipi}) 
for the positive and negative candidates.
The spectra were fitted with Gaussian signals
and linear backgrounds. The widths of the Gaussian functions were
constrained to be the same.
Reflections from $D^\pm \rightarrow K^\mp \pi^\pm \pi^\pm$ appeared near the 
$D_s^\pm \rightarrow K^\mp K^\pm \pi^\pm$
and $K^*(892) K^\pm$  signals, and as a broad shoulder below 
the $D^\pm \rightarrow \pi^\mp \pi^\pm \pi^\pm$ 
signal. These reflection regions were excluded from the fits.

The mass spectra for all $K^\mp K^\pm \pi^\pm$  combinations that survived 
the optimized cuts are shown in Fig.~\ref{fig_kkpi}.
The simultaneous fits of the $K^- K^+ \pi^+$ and $K^+ K^- \pi^-$ mass spectra
yielded  $1031 \pm 44$ and $1265 \pm 48$
signal events for the $D^+$ and $D^-$, respectively. 

To reconstruct the $D^\pm$ in the decay modes 
$\phi \pi^\pm$ with $\phi \rightarrow K^+ K^-$,
we required candidates to have 
$M(K^+ K^-)$ within $\pm 6$ MeV/$c^2$ of the $\phi$ mass.
Because of the real $K^+K^-$ background under the $\phi$ signal,
we performed $\phi$ mass sideband subtraction to determine 
the true number of candidates decaying to
$\phi \pi^\pm$. The $\phi$ sideband regions, 
which were chosen to have summed area
equal to the background in the signal region, 
were $0.990 < M(K^+K^-) < 1.000$~GeV/$c^2$ and  
$1.040 < M(K^+K^-) < 1.044$~GeV/$c^2$.
The simultaneous fits of the $\phi \pi^+$ and $\phi \pi^-$ mass spectra, 
shown in Fig.~\ref{fig_phi}, yielded $474 \pm 25$ and $598 \pm 28$
signal events for the $D^+$ and $D^-$, respectively. 

To reconstruct $D^\pm \rightarrow K^*(892) K^\pm$ 
with $\overline{K}^*(892)^0 \rightarrow K^- \pi^+$ and 
$K^*(892)^0 \rightarrow K^+ \pi^-$,
we required candidates to have 
$M(K^\mp \pi^\pm)$ within $\pm 45$ MeV/$c^2$ of the $K^*(892)$ mass.
Here, we also performed $K^*(892)$ mass sideband subtraction to determine 
the true number of candidates decaying to
$K^*(892) K^\pm$. The $K^*(892)$ sideband regions were
$0.680 < M(K^\mp \pi^\pm) < 0.750$~GeV/$c^2$ and 
$1.040 < M(K^\mp \pi^\pm) < 1.070$~GeV/$c^2$.
The simultaneous fits of the $\overline{K}^*(892)^0 K^+$ and 
$K^*(892)^0 K^-$ mass spectra, shown in 
Fig.~\ref{fig_kst}, yielded $239 \pm 18$ and $291 \pm 19$
signal events for the $D^+$ and $D^-$, respectively. 

In the study of the decay $D^\pm \rightarrow \pi^\mp \pi^\pm \pi^\pm$,
we investigated possible reflections from 
$D^\pm \rightarrow K^\mp \pi^\pm \pi^\pm$ and $K^\mp K^\pm \pi^\pm$,
and from random combinations of  
$D^0(\overline{D}^0) \rightarrow K^\mp \pi^\pm$, $\pi^\mp \pi^\pm$, and 
$K^\mp K^\pm$ with a $\pi^\pm$. The only significant reflection
was from $D^\pm \rightarrow K^\mp \pi^\pm \pi^\pm$. 
We excluded this region 
from the fits, as shown in Fig.~\ref{fig_3pi}.
The $\pi^- \pi^+ \pi^+$ and $\pi^+ \pi^- \pi^-$ mass spectra were 
simultaneously fit using four Gaussian functions for 
the $D^\pm$ and $D_s^\pm \rightarrow \pi^\mp \pi^\pm \pi^\pm$ signals.
The fits yielded $697 \pm 42$ and $851 \pm 48$ 
signal events for the $D^+$ and $D^-$, respectively. 

We determined the yields of the CF normalizing decay
mode $D^\pm \rightarrow K^\mp \pi^\pm \pi^\pm$ that survived 
the optimized cuts for each of the SCS decays 
$D^\pm \rightarrow K^\mp K^\pm \pi^\pm$, $\phi \pi^\pm$, $K^*(892) K^\pm$, 
and $\pi^\mp \pi^\pm \pi^\pm$.
The simultaneous fits of the $K^- \pi^+ \pi^+$  and $K^+ \pi^- \pi^-$ 
mass spectra with cuts 
optimized for $D^\pm \rightarrow K^\mp K^\pm \pi^\pm$ are shown in 
Fig.~\ref{fig_kpipi}.
The fits yielded $23465 \pm 183$ and $28014 \pm 201$ signal events for 
the $D^+$ and $D^-$, respectively. 
Fitting the  $K^- \pi^+ \pi^+$  and $K^+ \pi^- \pi^-$ mass
spectra with cuts optimized for $D^\pm \rightarrow \phi \pi^\pm$, 
$K^*(892) K^\pm$, and $\pi^\mp \pi^\pm \pi^\pm$ yielded 
$31084 \pm 254$,
$23765 \pm 182$, and
$20105 \pm 156$ signal events for the $D^+$, respectively.
The signal yields for the $D^-$ were
$37057 \pm 281$,
$28384 \pm 199$, and
$23744 \pm 169$, respectively.
We investigated different techniques for extracting the number of $D^+$
and $D^-$ signal events from the data of Fig.~\ref{fig_kpipi}, with special 
attention given to the excess of events seen just below the $D$ mass.
In the most extreme case, the ratio of $D^+$ to $D^-$ increased by
1.9\%, which changed the measured asymmetry by 0.9\%.
This variation is sufficiently small that
we neglected it in calculating the 90\% confidence level limits.

Using the measured yields of the various decay modes, we have calculated
the CP-violation asymmetry parameters $A_{CP}$
directly from Eqs.~(\ref{eq_2}) and~(\ref{eq_3}).
The central values have statistical errors of the order of 3--5\%
and are all consistent with zero asymmetry.
These results for $A_{CP}$  and the 90\% confidence level limits 
are summarized in Table~\ref{tab_1}. 

As discussed above, in order to extract $A_{CP}$ with no bias,
Eq.~(\ref{eq_eff}) must be satisfied.
Since we are normalizing to the CF decay 
$D^\pm \rightarrow K^\mp \pi^\pm \pi^\pm$, most sources 
of systematic error cancel. Detector asymmetries may 
lead to a fake CP asymmetry. However, we expect such 
effects to be momentum dependent.
We investigated any momentum dependence in our efficiencies by 
comparing the ratio of ratios of the yields of $D^+$ to $D^-$ for all 
decays, as they appear in the efficiency relation given above in 
Eq.~(\ref{eq_eff}), in bins of the $D^\pm$ momentum.
We also investigated momentum dependence in the asymmetry parameter 
itself by measuring $A_{CP}$ for the different decay modes in bins  
of $D^\pm$ momentum. Within our statistical errors, we saw no momentum 
dependence in either test. 
Thus, we do not assign any systematic errors for these sources.

Various additional sources of systematic uncertainty were investigated
by varying the cuts for the different decay modes,
the $\phi$ and $K^*(892)$ mass sideband regions, and the fitting procedure, 
and by performing reflection subtractions instead of excluding the 
reflection regions from the fits.
For all decay modes, the total systematic variations in $A_{CP}$ were 
found to be small compared to the statistical errors (less than half) 
and were thus neglected.

In conclusion, we have measured $A_{CP}$
for $D^\pm \rightarrow K^\mp K^\pm \pi^\pm$, $\phi \pi^\pm$, 
$K^*(892) K^\pm$, and $\pi^\mp \pi^\pm \pi^\pm$.  
We see no evidence of anomalous CP violation in SCS charm decay.
The statistical errors on our $A_{CP}$ measurements for the 
$D^\pm \rightarrow K^\mp K^\pm \pi^\pm$, $\phi \pi^\pm$,  
and $K^*(892) K^\pm$ are more than a factor of two smaller 
than the only previous measurements~\cite{ref_e687}, 
and we have made the 
first measurement of $A_{CP}$ for the decay 
$D^\pm \rightarrow \pi^\mp \pi^\pm \pi^\pm$.

%%%%%%

\begin{ack}

We gratefully acknowledge the assistance of the staffs of Fermilab 
and of all the
participating institutions. This research was supported by the 
Brazilian Conselho Nacional de Desenvolvimento Cient\'{i}fico e 
Technol\'{o}gio, CONACyT (Mexico), the Israeli Academy of Sciences
and Humanities, the U.S. Department of Energy, the U.S.-Israel
Binational Science Foundation and the U.S. National Science 
Foundation. Fermilab is operated by the Universities Research 
Association, Inc., under contract with the United States Department
of Energy.

\end{ack}

%%%%%%

%%%%%%

%%%%%%

\begin{table}
\caption{Summary of our measured CP-violation asymmetry parameters
$A_{CP}$ and the 90\%
confidence level limits for the singly Cabibbo-suppressed decays 
$D^+ \rightarrow K^- K^+ \pi^+$, $\phi \pi^+$, 
$\overline{K}^*(892)^0 K^+$, and $\pi^- \pi^+ \pi^+$.}
\label{tab_1}
\bigskip
\begin{center}
\begin{tabular}{l c r}
\hline
Decay Mode & $A_{CP}$ & 90\% CL Limits (\%) \\
\hline
$K K \pi$   & $-0.014 \pm 0.029$ & $-6.2 < A_{CP} < 3.4$ \\ 
$\phi \pi$  & $-0.028 \pm 0.036$ & $-8.7 < A_{CP} < 3.1$ \\
$K^*(892) K$     & $-0.010 \pm 0.050$ & $-9.2 < A_{CP} < 7.2$ \\
$\pi\pi\pi$ & $-0.017 \pm 0.042$ & $-8.6 < A_{CP} < 5.2$ \\
\hline
\end{tabular} 
\end{center} 
\end{table}

%%%%%%

\end{document}